# Coarse-grained entropy rates for characterization of complex time series

Milan Paluš

*Santa Fe Institute, 1399 Hyde Park Road, Santa Fe, NM 87501, USA; and*

*Institute of Computer Science, Academy of Sciences of the Czech Republic*

*Pod vodárenskou věží 2, 182 07 Prague 8, Czech Republic*

*E-mail:* `mp@santafe.edu, mp@uivt.cas.cz`

November 23, 1995

**Abstract**

A method for classification of complex time series using coarse-grained entropy rates (CER's) is presented. The CER's, which are computed from information-theoretic functionals – redundancies, are relative measures of regularity and predictability, and for data generated by dynamical systems they are related to Kolmogorov-Sinai entropy. A deterministic dynamical origin of the data under study, however, is not a necessary condition for the use of the CER's, since the entropy rates can be defined for stochastic processes as well. Sensitivity of the CER's to changes in data dynamics and their robustness with respect to noise are tested by using numerically generated time series resulted from both deterministic – chaotic and stochastic processes. Potential application of the CER's in analysis of physiological signals or other complex time series is demonstrated by using examples from pharmaco-EEG and tremor classification.

## 1 Introduction

A number of descriptive measures, like dimensions, entropies and Lyapunov exponents, for characterization of complex time series have been developed, based on concepts from nonlinear dynamics and theory of deterministic chaos [1, 2, 3, 4]. These measures have well-defined meanings when analyzed data have been indeed generated by a low-dimensional deterministic system. Analyzing experimental time series, like those in biology and medicine, underlying dynamical mechanisms are usually unknown, and, due to questionable reliability of dimensional or Lyapunov exponents algorithms applied to short and noisy data, results eventually supporting the hypothesis of low-dimensional chaos cannot be taken without reservations. Some authors [5, 6, 7, 8] do not insist any more on interpretation of their results, like finite dimension estimates, as evidence for underlying low-dimensional chaos, but propose these measures, mostly the correlation dimension (CD) [1, 2], as measures for relative characterization of different datasets. In the case of physiological time series, recorded in different physiological states, these measures are proposed to characterize physiological states of an organism or its parts.

Although these authors demonstrate that such dimensional estimates can have some discriminating power with respect to time series recorded in different experimental conditions, considering the underlying processes can be high-dimensional or stochastic, these low numbers, formally obtained from dimensional algorithms, are probably spurious and have no theoretically justified meaning and interpretation. Moreover, it can hardly be established, how robust with respect to noise, or how sensitive to changes in underlying dynamics the measures like the CD are, when applied to relative characterization of processes, which dimensionality can be effectively infinite.

In this paper we propose alternative measures, called coarse-grained entropy rates (CER's), which are easy to compute and applications of which do not involve the above theoretical and practical problems. For data generated by chaotic dynamical systems the CER's are related to Kolmogorov-Sinai entropy. A deterministic dynamical origin of the data under study, however, is not a condition necessary for the use of the CER's, since the entropy rates can be defined for stochastic processes as well. We argue, however, that the exact entropy rates cannot be computed in majority of experimental situations. Therefore we do not estimate the limit values given in the definitions of the exact entropy rates, we rather propose coarse-grained quantities, which are related to the exact entropy rates, however, their values can depend on particular experimental and numerical conditions. Thus the CER's are not meant as absolute quantities able to classify systems in general, or to identify chaotic systems, but rather as relative quantities for comparison of datasets recorded in the same experimental conditions and processed using the same numerical procedures. On the other hand, the CER's have the same theoretical interpretation as the exact entropy rates: The CER's are (relative) measures of regularity and predictability, i.e., if one dataset gives higher CER than the other, the former is more irregular and less predictable than the latter, and, as we demonstrate below, the exact entropy rates (or the Kolmogorov-Sinai entropies) of the underlying processes are in the same relation.

The CER's, proposed in this paper, are computed from information-theoretic functionals called marginal redundancies, which, together with the exact entropy rates are briefly introduced in Sec. 2. Further details can be found in Refs. [11, 15, 17] and references therein. For deeper understanding of the theoretical background we recommend Refs. [9, 10, 12, 13, 14]. In Section 3 we define the coarse-grained entropy rates. Their numerical properties, sensitivity to changes in dynamics underlying analyzed data, robustness with respect to additive noise and some transformations of data are studied in Sec. 4. Potential applications of the CER's in analysis of physiological signals or other complex time series are demonstrated in Sec. 5 by using examples from pharmaco-EEG and tremor classification. Conclusion is given in Sec. 6.

## 2 Marginal redundancies and entropy rates

Consider $n$ discrete random variables $X_1, \ldots, X_n$ with sets of values $\Xi_1, \ldots, \Xi_n$, respectively. The probability distribution for an individual $X_i$ is $p(x_i) = \Pr\{X_i = x_i\}$, $x_i \in \Xi_i$. We denote the probability distribution function by $p(x_i)$, rather than $p_{X_i}(x_i)$, for convenience. Analogously, the joint distribution for the $n$ variables $X_1, \ldots, X_n$ is $p(x_1, \ldots, x_n) = \Pr\{(X_1, \ldots, X_n) = (x_1, \ldots, x_n)\}$, $(x_1, \ldots, x_n) \in \Xi_1 \times \ldots \times \Xi_n$.

The marginal redundancy $\varrho(X_1, \ldots, X_{n-1}; X_n)$, in the case of two variables also known as mutual information $I(X_1; X_2)$, quantifies the average amount of information about the variable $X_n$, contained in the $n-1$ variables $X_1, \ldots, X_{n-1}$, and is defined as:

$$\varrho(X_1, \ldots, X_{n-1}; X_n) = \sum_{x_1 \in \Xi_1} \ldots \sum_{x_n \in \Xi_n} p(x_1, \ldots, x_n) \log \frac{p(x_1, \ldots, x_n)}{p(x_1, \ldots, x_{n-1})p(x_n)}. \tag{1}$$

Now, let $\{X_i\}$ be a stochastic process, i.e., an indexed sequence of random variables, characterized by the joint probability distribution function $p(x_1, \ldots, x_n)$. The entropy rate of $\{X_i\}$ is defined as

$$h = \lim_{n \to \infty} \frac{1}{n} H(X_1, \ldots, X_n), \tag{2}$$

where $H(X_1, \ldots, X_n)$ is the joint entropy of the $n$ variables $X_1, \ldots, X_n$ with the joint distribution $p(x_1, \ldots, x_n)$:

$$H(X_1, \ldots, X_n) = -\sum_{x_1 \in \Xi_1} \ldots \sum_{x_n \in \Xi_n} p(x_1, \ldots, x_n) \log p(x_1, \ldots, x_n). \tag{3}$$

A way from the entropy rate of a stochastic process to the Kolmogorov-Sinai entropy (KSE) of a dynamical system can be straightforward due to the fact that any stationary stochastic process correspond to a measure-preserving dynamical system, and vice versa [13]. Then for the definition of the KSE we can consider the equation (2), however, the variables $X_i$ should be understood as $m$-dimensional variables, according to a dimensionality of a dynamical system. If the dynamical system is evolving on continuous (probability) measure space, then any entropy depends on a partition $\xi$ chosen to discretize the space and the KSE is defined as a supremum over all finite partitions [12, 13, 14].

Possibilities to compute the entropy rates from data are limited to a few exceptional cases: for stochastic processes it is possible, e.g., for finite-state Markov chains [9]. In the case of a dynamical system on continuous measure space the KSE can be, in principle, reliably estimated, if the system is low-dimensional and a large amount of (practically noise-free) data is available. In such a case, Fraser [11] proposed to estimate the KSE of a dynamical system from the asymptotic behavior of the marginal redundancy, computed from a time series generated by the dynamical system. In such an application one deals with a time series $\{y(t)\}$, considered as a realization of a stationary and ergodic stochastic process $\{Y(t)\}$. Then, due to ergodicity, the marginal redundancy (1) can be estimated using time averages instead of ensemble averages, and, the variables $X_i$ are substituted as

$$X_i = y(t + (i-1)\tau). \qquad (4)$$

Due to stationarity the marginal redundancy

$$\varrho^n(\tau) \equiv \varrho(y(t), y(t+\tau), \ldots, y(t+(n-2)\tau); y(t+(n-1)\tau)) \qquad (5)$$

is a function of $n$ and $\tau$, independent of $t$.

It was shown [11, 15], that if the underlying dynamical system is $m$-dimensional and the marginal redundancy $\varrho^n(\tau)$ is estimated using a partition fine enough (to attain so-called generating partition [11, 12, 14]), then the asymptotic behavior

$$\varrho^n(\tau) \approx H_1 - |\tau|h \qquad (6)$$

is attained for $n = m+1, m+2, \ldots$, for some range of $\tau$. The constant $H_1$ is related to $\varrho^n(0)$.

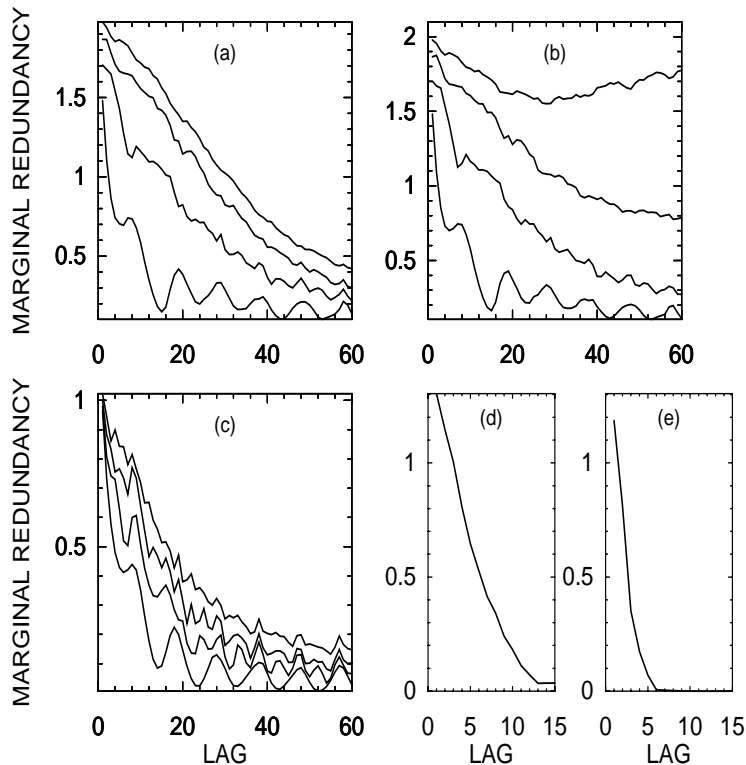

Figure 1: (a–c) Marginal redundancy as function of time lag $\tau$ for a time series generated by the chaotic Lorenz system: (a) $N = 1$ million samples, $Q = 16$, (b) $N = 16,384$, $Q = 16$, and (c) $N = 16,384$, $Q = 5$. The four different curves are marginal redundancies for different embedding dimensions, $n = 2 - 5$, reading from bottom to top. (d, e) Marginal redundancy vs. time lag for two states with different Kolmogorov-Sinai entropies of the chaotic baker map, $N = 131,072$ samples, $Q = 8$, embedding dimension $n = 2$. The data were generated using the parameter $\alpha = 0.1$ (d) and $\alpha = 0.3$ (e), see also Fig. 3a.

The marginal redundancies $\varrho^n(\tau)$, $n = 2 - 5$, for the chaotic Lorenz system [16] are presented in Fig. 1a. The Lorenz system is three-dimensional and $\varrho^n(\tau)$ for $n = 4$ and $5$ and lags $\tau = 5 - 40$ (approximately) is close to a linearly decreasing function so that the KSE $h$ can be estimated as its slope according to (6).

To obtain such a result, however, a relatively fine partition and an adequate amount of data must be used. In the case, presented in Fig. 1a, the time series length $N$ was one million samples and the partition was based on $Q = 16$ equiquantal marginal boxes.[1] If the same partition ($Q = 16$) is used for shorter time series, $N = 16,384$, the results are distorted (Fig. 1b). The reasons of this distortion are discussed in [15], where also the following requirement is proposed for the effective[2] series length $N$, necessary for the estimation of the $n$-dimensional redundancy using $Q$ equiquantal marginal boxes:

$$N \geq Q^{n+1}, \tag{7}$$

otherwise the results are distorted as in the example in Fig. 1b. The adequate partition ($Q = 5$ for $N = 16,384$, Fig. 1c) is not fine enough to attain the asymptotic behavior (6) and no linearly decreasing region in $\varrho^n(\tau)$ as a function of $\tau$ is detected. This means, that having a limited amount of data, the KSE cannot be estimated even approximately. Similar restrictions can be found also for different methods for estimation of the KSE or the Lyapunov exponents [18].

## 3 Coarse-grained entropy rates

In experimental practice the analyzed time series are usually short and contaminated by noise, so even if they resulted from low-dimensional chaotic processes, estimation of their KSE is practically impossible. And in many experiments actual dynamical mechanisms, underlying analyzed data, are unknown, and can be either high-dimensional deterministic or stochastic. As we pointed above, unlike the dimensions or Lyapunov exponents, the entropy rates are meaningful quantities for characterization of stationary processes irrespectively of their origin. The problem is, however, that the exact entropy rate of a process usually cannot be estimated. In order to utilize the concept of the entropy rates in time-series analysis we propose to give up the effort for estimating the exact entropy rates, and to define "coarse-grained entropy rates" (CER's) instead. The CER's are not meant as estimates of the exact entropy rates, but as quantities which can depend on a particular experimental and numerical set-up, however, quantities which have the same meaning as the exact entropy rates, i.e., which can be used as measures of regularity and predictability of analyzed time series in the relative sense: Two or several datasets can be compared according to their regularity and predictability, providing they were measured in the same experimental conditions and their CER's were estimated using the same numerical parameters, defined and discussed below.

The CER's are coarse-grained in space and their estimation is limited to finite time interval:
1. Space: If a time series resulted from a process evolving on a continuous[3] measure space, a partition used in estimating the CER is only as fine as a series length allows (eq. 7).
2. Time: The limit for $n \to \infty$ is not taken into account and used $n$ again depends on the series length. In many applications $n = 2$ or 3 is sufficient. Also the range of $\tau$ is limited (see the definitions below).

The most straightforward definition of a CER can be based on (6):

$$h^{(0)} = \frac{\varrho^n(\tau_0) - \varrho^n(\tau_1)}{\tau_1 - \tau_0}. \tag{8}$$

This definition[4], further referred to as the CER $h^{(0)}$, can be heavily influenced by the choice of $\tau_0$ and $\tau_1$, as far as formula (8) is directly related to the method of estimating the KSE from the linearly decreas-ing marginal redundancies $\varrho^n(\tau)$ (Fig. 1a), which, as we argued above, cannot be obtained in majority of experimental applications. Estimating $\varrho^n(\tau)$ from an experimental time series, which is short and/or high-dimensional or stochastic, the marginal redundancies $\varrho^n(\tau)$ do not decrease linearly, but in an exponential or power-law way (Fig. 1c), or even not monotonically – Figs. 5e,f present $\varrho^n(\tau)$ of a human electroencephalogram (EEG), in which a long-term decrease is modulated by faster (about 10 Hz) oscillations. Thus the CER $h^{(0)}$ depends on the choice of $\tau_0$ and $\tau_1$ and there is no criterion how to find "the best" $\tau$'s.

---

[1] We estimate the redundancies by the box-counting method adaptive in one dimension, i.e., the marginal boxes are defined in such a way that there is approximately the same number of points in each marginal box. Thus a partition is defined by a number $Q$ of the equiquantal marginal boxes. For the embedding dimension $n$ the total number of the partition boxes is $Q^n$. For more details see [15, 17, 19].

[2] The effective series length $N$ is $N = N_0 - (n-1)\tau$, where $N_0$ is the total series length, $n$ is the embedding dimension and $\tau$ is the time delay used in the estimation of $\varrho^n(\tau)$.

[3] Of course, any digitized experimental time series is discrete, however, a marginal partition given by a standard equipment is usually too fine and must be coarsened. For instance, according to eq. (7) the necessary series length for estimation of the 2-dimensional redundancy using 12-bit precision (4096 bins) is $68 \times 10^9$ samples.

[4] For the particular choice $\tau_0 = 0$ this definition is related to the approximate entropy ApEn introduced by Pincus et al. [20], however, the ApEn is estimated from correlation integrals [21], while the CER $h^{(0)}$ is computed from the marginal redundancy estimated by an adaptive box-counting method – see Footnote 1. One can also estimate the (generalized) redundancies using the correlation integrals [22], however, we have not explored this possibility yet.

For an alternative definition of the CER we consider the following properties of $\varrho^n(\tau)$:

- For a process with a positive entropy rate the marginal redundancy $\varrho^n(\tau) \to 0$ for $\tau \to \infty$. In finite-precision computation, the (coarse-grained) marginal redundancy of such a process decreases to zero value in finite $\tau$, and the integral $\int \varrho^n(\tau) d\tau$ is finite.

- The integral $\int \varrho^n(\tau) d\tau$ (the area under the curve) depends on a particular entropy rate of a process under study, as demonstrated in Figs. 1d and 1e, where we present $\varrho^2(\tau)$ for two states with different KSE's of the chaotic baker map (discussed in detail in Sec. 4).

Then, in a particular application, we compute the marginal redundancies $\varrho^n(\tau)$ for all analyzed datasets and find such $\tau_{max}$ that for $\tau' \geq \tau_{max}$: $\varrho^n(\tau') \approx 0$ for all the datasets. Then we define a norm of the marginal redundancy

$$||\varrho^n|| = \frac{\sum_{\tau=\tau_0}^{\tau_{max}} \varrho^n(\tau)}{\tau_{max} - \tau_0}. \qquad (9)$$

In experimental applications the lags $\tau$ are discrete and thus the integral $\int \varrho^n(\tau) d\tau$ was substituted by the sum in (9). The lag $\tau_0$ is usually set to zero.

Having defined the norm $||\varrho^n||$, the difference $\varrho^n(\tau_0) - ||\varrho^n||$ can be considered as the alternative definition of the CER. We have found, however, that the definition of the CER, which does not depend on absolute values of $\varrho^n(\tau)$, has better numerical properties, namely the estimates are more stable and less influenced by noise. Thus, we define the CER $h^{(1)}$ as

$$h^{(1)} = \frac{\varrho^n(\tau_0) - ||\varrho^n||}{||\varrho^n||}. \qquad (10)$$

## 4 Properties of CER's – numerical examples

Consider an autoregressive process (ARP) given as

$$y_t = c \sum_{k=1}^{10} a_k y_{t-k} + \sigma e_t, \qquad (11)$$

where $a_{k=1,\ldots,10} = 0, 0, 0, 0, 0, .19, .2, .2, .2, .2$, $\sigma = 0.01$ and $e_t$ are Gaussian deviates with zero mean and unit variance. For $c = 1$ this ARP has long coherence time [19], for $c < 1$ the coherence time decreases and the entropy rate increases. In particular, we can generate a number of the ARP's with different $c$'s and thus with different entropy rates. The entropy rates of such ARP's should monotonically decrease with increasing $c$. Figure 2 presents the coarse-grained entropy rates for 100 ARP's with $c$ increasing from 0.5 to 0.9. For the estimation of $h^{(0)}$ we set $\tau_0 = 0$ and $\tau_1 = 1$ (sample), for $h^{(1)}$ $\tau_0 = 0$ and $\tau_{max} = 100$ (samples) were set. The results in Figs. 2a,b were obtained using the series length $N = 16,384$ samples and $Q = 16$ equiquantal marginal levels, in Figs. 2c,d $N = 1,024$ and $Q = 8$ were used. The embedding dimension $n = 2$ was used.

The CER $h^{(1)}$, estimated from 16K samples and $Q = 16$ (Fig. 2b) exhibits the expected smooth monotonic decrease. Using shorter time series ($N = 1$K, $Q = 8$, Fig. 2d) the estimates are less stable, i.e., fluctuations from the smooth monotonic curve occur. The estimates of CER $h^{(0)}$ seems less stable than those of $h^{(1)}$ using the same $N$ and $Q$. (Cf. Figs. 2a and 2b for the 16K estimates and 2c and 2d for the 1K estimates.)

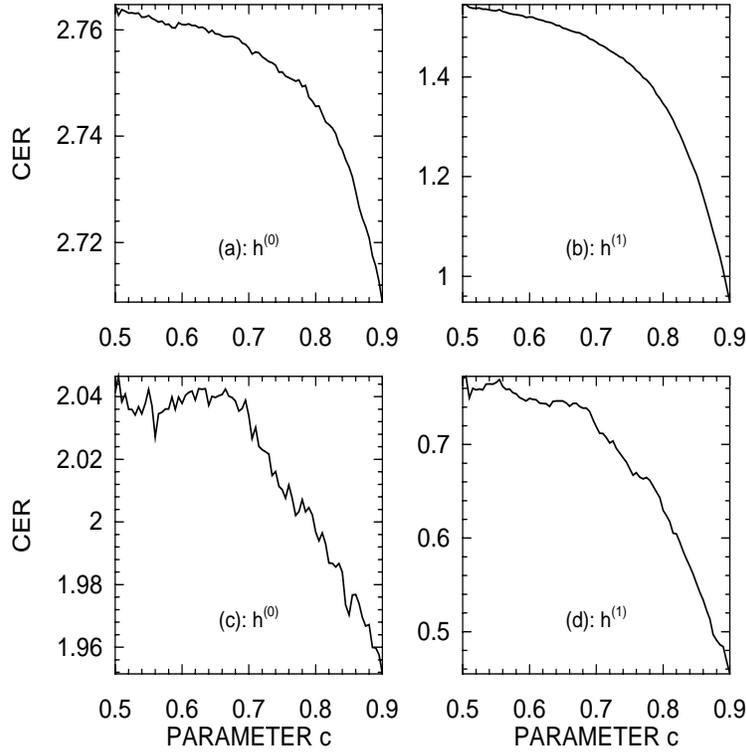

Figure 2: *Coarse-grained entropy rates* $h^{(0)}$, $\tau_0 = 0$, $\tau_1 = 1$ *(a,c), and* $h^{(1)}$, $\tau_0 = 0$, $\tau_{max} = 100$ *(b,d);* $N = 16,384$ *and* $Q = 16$ *(a,b),* $N = 1,024$ *and* $Q = 8$ *(c,d), as functions of the parameter c, computed from one hundred time series generated by the autoregressive process, which exact entropy rate is a smooth monotonically decreasing function of the parameter c. Embedding dimension* $n = 2$.

Another example for the study of behavior of the CER's is the chaotic baker transformation:

$$(x_{n+1}, y_{n+1}) = (\lambda x_n, \frac{1}{\alpha} y_n)$$

for $y_n \leq \alpha$, or:

$$(x_{n+1}, y_{n+1}) = (0.5 + \lambda x_n, \frac{1}{1-\alpha}(y_n - \alpha)) \tag{12}$$

for $y_n > \alpha$;
$0 \leq x_n, y_n \leq 1$, $0 < \alpha < 1$, $\lambda$ was set to $\lambda = 0.25$. For this system the positive Lyapunov exponent, or, equivalently, the Kolmogorov-Sinai entropy can be expressed analytically as the function of the parameter $\alpha$ [23, 24]:

$$h(\alpha) = \alpha \log \frac{1}{\alpha} + (1-\alpha) \log \frac{1}{1-\alpha}. \tag{13}$$

We can generate a number of chaotic time series with different positive Lyapunov exponents and compare the behavior of the CER's with the exact dependence of the positive Lyapunov exponent (or, equivalently, of the KSE) on the parameter $\alpha$, displayed in Fig. 3a. Figures 3b–g display the same dependence of the CER's $h^{(0)}$ (Figs. 3b,d,f) and $h^{(1)}$ (Figs. 3c,e,g) estimated using different time series lengths $N$ and numbers $Q$ of the equiquantal marginal levels: $N = 16K$ and $Q = 16$ (Figs. 3b,c), $N = 1K$ and $Q = 8$ (Figs. 3d,e) and $N = 256$ and $Q = 4$ (Figs. 3f,g). The embedding dimension used was $n = 2$, the lags $\tau_0 = 0$, $\tau_1 = 1$ and $\tau_{max} = 100$.

The CER $h^{(1)}$ for $N = 16K$ and $Q = 16$ (Fig. 3c) very well mimics the dependence of the positive Lyapunov exponent on the parameter $\alpha$. Similarly, like in the above case of the ARP, the results obtained from shorter time series are less stable for both the $h^{(0)}$ and $h^{(1)}$, while for longer time series the CER $h^{(1)}$ seems to perform better comparison of different time series than the CER $h^{(0)}$, namely the fluctuations from the expected curve are larger in the case of $h^{(0)}$.

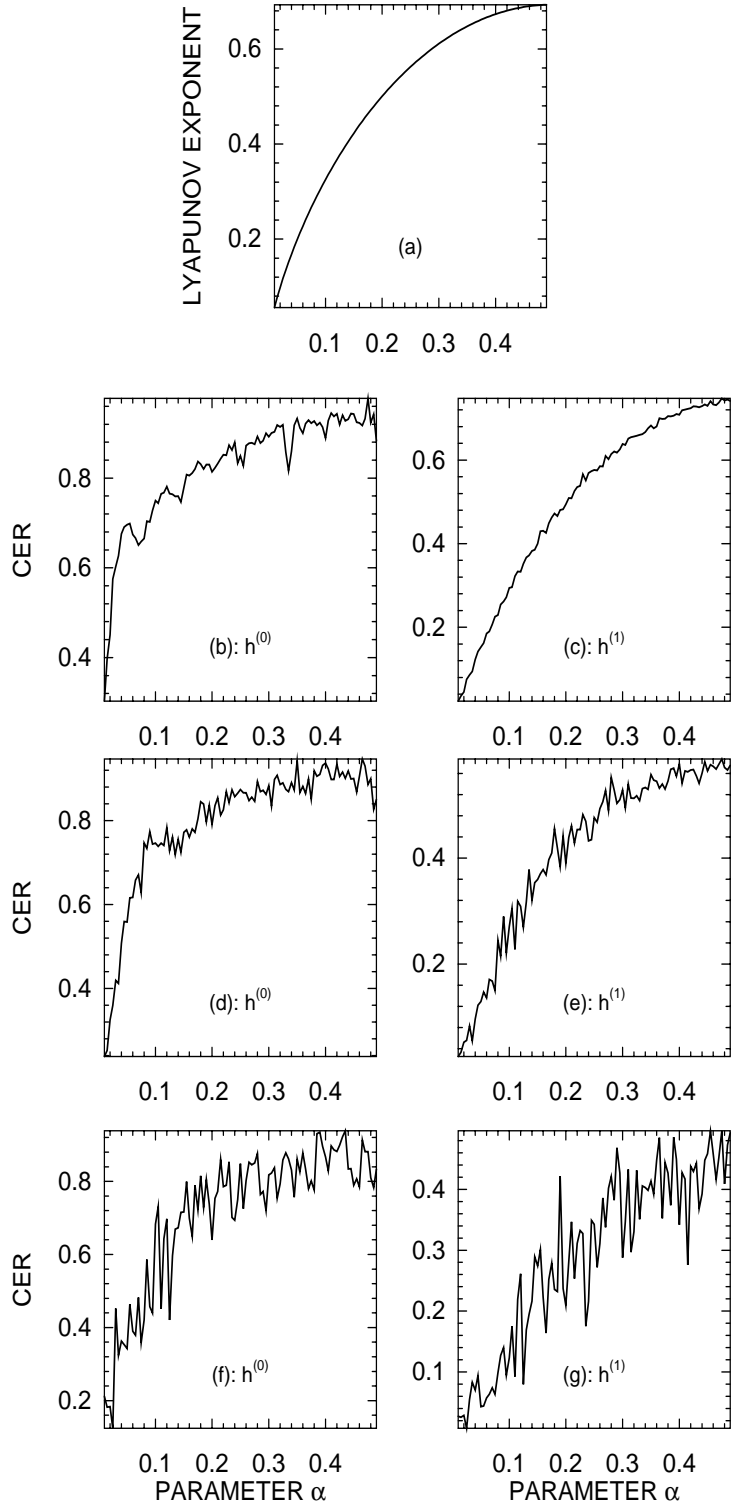

Figure 3: (a) Positive Lyapunov exponent (or Kolmogorov-Sinai entropy) of the chaotic baker map computed as the analytical function (13) of the parameter $\alpha$. (b–g) Coarse-grained entropy rates $h^{(0)}$, $\tau_0 = 0$, $\tau_1 = 1$ (b,d,f), and $h^{(1)}$, $\tau_0 = 0$, $\tau_{max} = 100$ (c,e,g); $N = 16,384$ and $Q = 16$ (b,c), $N = 1,024$ and $Q = 8$ (d,e), $N = 256$ and $Q = 4$ (f,g), as functions of the parameter $\alpha$, computed from ninety-seven time series generated by the chaotic baker maps with the parameter $\alpha$ changing from 0.01 to 0.49. Embedding dimension $n = 2$.

The above examples demonstrate that the CER's $h^{(0)}$ and $h^{(1)}$ can distinguish time series with different exact entropy rates. The CER $h^{(0)}$ is measured in bits or nats per a time unit, the CER $h^{(1)}$ is a dimensionless quantity. As stressed above, neither $h^{(0)}$ nor $h^{(1)}$ are meant as estimates of exact entropy rates or Kolmogorov-Sinai entropy, but as measuring tools for relative comparison of different datasets.

The CER's, computed from the marginal redundancy, are invariant with respect to linear transformations of data, and, due to the way of estimating the marginal redundancy[5] neither smooth monotonous nonlinear transformations change the results.

In experimental practice data are usually contaminated by noise. Influence of additive Gaussian noise on the estimations of the CER's was also studied. The additive noise increases the values of the CER's, however, the relative comparison of different datasets is not changed. In other words, when the same numerical experiments as those presented in Figs. 3b,c were performed using noisy data, the same curves as those in Figs. 3b,c were obtained, just the values on the y-axis were different. In another numerical experiment the robustness of the CER's with respect to noise was compared with behavior of "fine-grained" measures – Lyapunov exponents. To have a "controlled" experiment, again the baker transformation (12) was used, however, in this case only three states of the system, with three different values of the parameter $\alpha$ (0.1, 0.2 and 0.3) were used. In each state ten independent "measurements" were made, i.e., for each of the above $\alpha$'s ten time series, 1024 samples long, were generated and a defined amounts of additive Gaussian noise were added to the generated data. Then the series were quantitatively characterized by:
a) computing the CER $h^{(1)}$ using $Q = 8$, $\tau_0 = 0$ and $\tau_{max} = 100$.
b) Estimating the Lyapunov exponents using a Jacobian based algorithm [25, 26] implemented in a program written by E. Kostelich [27]; embedding dimension $n = 2$ and time delay $\tau = 1$ sample were used, the larger Lyapunov exponent $\lambda_1$ was recorded.
c) Estimating the largest Lyapunov exponent using the direct method according to Wolf et al. [28], $n = 2$, $\tau = 1$ and 2 were used.
The Jacobian method for estimating the Lyapunov exponents was found very vulnerable to noise, in particular, with 50% of noise in the data (50% of noise means that the standard deviation of the noise was equal to 50% of the standard deviation of the original noise-free data) the estimated Lyapunov exponents totally failed to distinguish the three states of the baker system. The direct method was found more robust than the Jacobian based algorithm. The comparison of the results of the direct Lyapunov exponent algorithm and the CER's $h^{(1)}$ is presented in Fig. 4 (the Lyapunov exponents $\lambda_1$ – left panels, the CER's $h^{(1)}$ – right panels, squares represent means of the ten "measurements", standard deviations are marked by vertical lines). Applied to the noise-free data (upper panels) the direct Lyapunov algorithm provides better distinction (smaller relative variances) than the CER's. With 30% of the additive noise (percentages in the above defined sense, middle panels) the discriminating power of $\lambda_1$ is comparable with that of $h^{(1)}$, while having 50% of the noise in the data the Lyapunov exponents have larger relative variance and provide worse discrimination of the system states than the CER's $h^{(1)}$ (Fig. 4, lower panels)[6].

Like the Lyapunov exponents also other "fine-grained" measures are very vulnerable to noise. For instance, Kostelich & Yorke [29] demonstrated how additive noise affected estimations of the correlation dimension. Therefore the CER's, which are very robust with respect to noise, are more suitable for real-world applications than any of the chaos-based measures defined in terms of vanishing distance between points. And last but not the least argument — compare the computing times necessary to obtain the results in one of the six panels in Fig. 4 (30 values of either $h^{(1)}$ or $\lambda_1$): 38 seconds for the CER's $h^{(1)}$ and 793 seconds for the Lyapunov exponents $\lambda_1$, using the same SPARC-station IPX.

---

[5] The marginal "equiquantization", mentioned in Footnote 1, effectively means a transformation of data into a uniform marginal distribution. A smooth monotonous (invertible) nonlinear transformation of a time series means only a change of the marginal distribution, which is eliminated by the equiquantization.

[6] Actually, the distinction by $h^{(1)}$ seems better for noisy data than for noise-free data. It is caused by the fact that the variances of all presented estimates of $h^{(1)}$ are similar, while the values of $h^{(1)}$ increase with increasing the amount of noise, and the relative variances decrease.

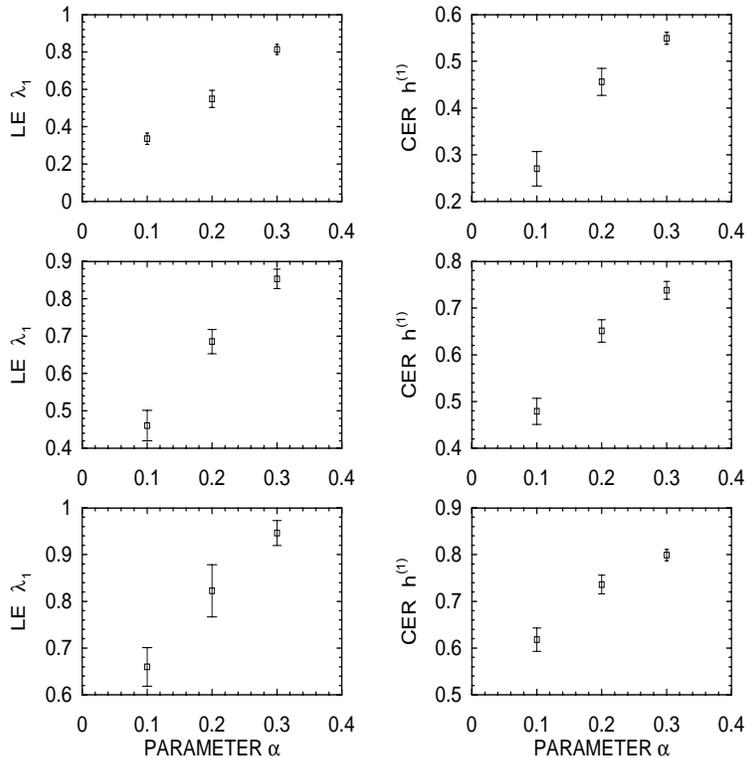

Figure 4: *Positive Lyapunov exponents $\lambda_1$, estimated by the direct method (left-side panels) and CER's $h^{(1)}$ (right-side panels), estimated from the baker series with different amounts of additive Gaussian noise. The squares and the vertical lines depict means and standard deviations, respectively, of the groups of ten independent realizations of 1024-sample series that were generated for each of the three values (0.1, 0.2 and 0.3) of the parameter $\alpha$. The results for noise-free data are presented in the upper panels, the middle panels present results for 30%, and the lower panels for 50% (relative to SD of noise-free data) of noise added to the data.*

## 5 Applications of CER's – characterization of pharmaco-EEG's and tremors

In practical applications the choice between $h^{(0)}$ and $h^{(1)}$ depends on an available amount of data. As far as we have demonstrated that, for longer time series, the CER $h^{(1)}$ have better numerical properties than $h^{(0)}$ and, unlike $h^{(0)}$, $h^{(1)}$ is independent of the choice of the time lag $\tau_1$ ($\tau_0$ is usually set to 0, $\tau_{max}$ is given by the data), we prefer the application of the CER $h^{(1)}$, if the amount of available data is sufficient for estimating the marginal redundancy $\varrho^n(\tau)$ for the range of the lags $\tau$ large enough to attain the lag $\tau_{max}$, $\varrho^n(\tau \geq \tau_{max}) \approx 0$, at least for $n = 2$. Analyzing short time series, when $\tau_{max}$ is comparable to the series length and computing $\varrho^n(\tau)$ for a large range of $\tau$'s can drastically decrease the effective series length (see Footnote 2) the CER $h^{(0)}$ can be the better choice: computing $h^{(0)}$ with $n = 2$, $\tau_0 = 0$ and $\tau_1 = 1$, the effective series length is $N_{total} - 1$, so that the maximum available effective series length is used to secure maximum available reliability and stability of the results.

In spite of invariance of the CER's with respect to linear and some nonlinear transformations (as discussed above), we propose to apply the CER's as relative measures for comparison of datasets recorded in the same experimental conditions and processed using the same numerical parameters. A demonstrative example can be data from a pharmaco-EEG[7] study.

---

[7] Pharmacoelectroencephalography (pharmaco-EEG) is a neuroscientific discipline oriented to electrophysiological brain-

The electroencephalogram (EEG) of a healthy human volunteer in a state of relaxed vigilance with closed eyes was recorded before and several times after (0.5, 1, 2, 3, 4 and 6 hours) a dose of alcohol (50 ml of ethanol mixed with a soft drink) was administered. Concentration of ethanol in blood (Fig. 5a) was measured from breath in the same intervals as the EEG was recorded. Reported results were obtained from the EEG signal recorded with the sampling frequency 128 Hz in position $O_1$ with the Goldman reference electrode.

Having a large amount of data (16,384 samples for each recording) available, we chose the CER $h^{(1)}$. The marginal redundancies $\varrho^n(\tau)$ decrease to zero in lags shorter than 0.5 sec. and $\tau_{max} = 50$ samples was chosen for the estimation of $h^{(1)}$, embedding dimension $n = 3$, and $Q = 4$ marginal equiquantal bins were used. The result – dependence of the CER $h^{(1)}$ on the time after the dosage of alcohol – is presented in Fig. 5b: Increase of the concentration of alcohol in blood induces decrease of the coarse-grained entropy rate $h^{(1)}$. Figures 5c and 5d present the results of standard spectral analysis, namely the spectral powers in the alpha band (8 − 13 Hz) and in the beta band (13 - 32 Hz). Presence of alcohol in blood causes increase of the alpha activity and decrease of the beta activity in the EEG of this volunteer. Particularly, the spectral power in the beta band very well correlates with the CER $h^{(1)}$. The CER reflects physiologically meaningful information which, however, can also be obtained from the spectral analysis. In order to understand this result we will look directly at the related marginal redundancies $\varrho^3(\tau)$, computed from the EEG data and also from so called isospectral surrogate data.

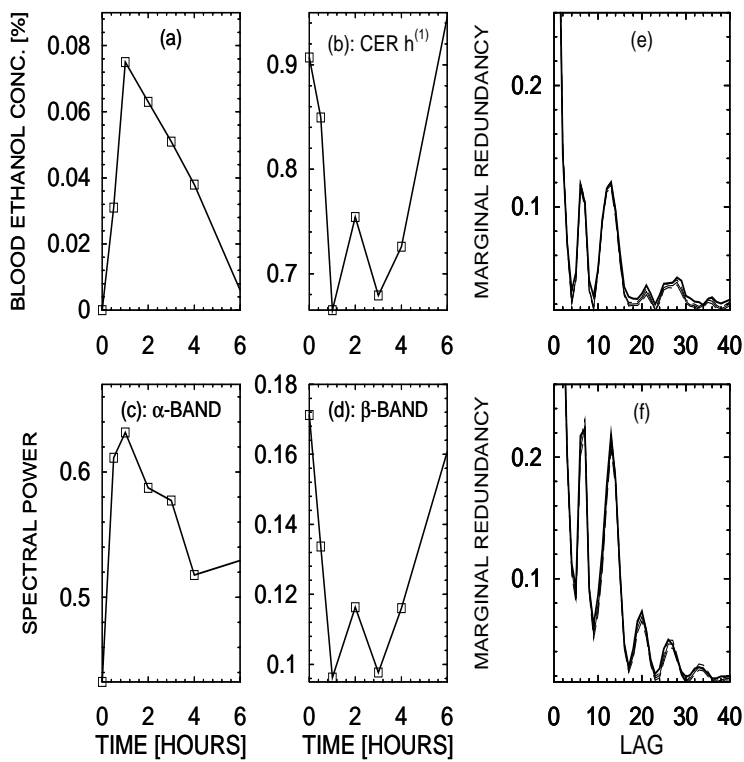

Figure 5: (a) Concentration of alcohol in blood, (b) coarse-grained entropy rate $h^{(1)}$ ($N = 16,384$, $Q = 4$, $n = 3$, $\tau_0 = 0$, $\tau_{max} = 50$), (c) spectral power in the alpha band (8 − 13 Hz), (d) spectral power in the beta band (13 − 32 Hz), as functions of time after a dose of alcohol to a healthy volunteer. The CER $h^{(1)}$ and the spectral parameters were obtained from the EEG signal recorded in the position $O_1$, Goldman average reference electrode, sampling rate 128 Hz. (e, f) Marginal redundancies for the EEG recorded in time 0 hours (e) and 1 hour (f) after the alcohol administration, and the marginal redundancies for corresponding isospectral surrogate data (thinner but coinciding lines), $N = 16,384$, $Q = 4$, $n = 3$, lags 1 − 40 samples.

---

research in pharmacology, clinical pharmacology, neurotoxicology, pharmacopsychiatry and related fields, in which effects of psychoactive substances on the EEG are studied and evaluated [30].

The *surrogate data* have been introduced as a tool for detecting nonlinearity in time series [31]. The surrogate-data based nonlinearity tests usually consist of computing a *nonlinear* statistic from data under study and from an ensemble of the surrogates – realizations of a linear stochastic process, which mimics "linear properties" of the studied data. If the computed statistic for the original data is significantly different from the values obtained for the surrogate set, one can infer that the data were not generated by a linear process; otherwise the null hypothesis, that a linear model fully explains the data, is accepted. For the purpose of such tests the surrogate data must preserve the spectrum[8] and consequently, the autocorrelation function of the series under study. An isospectral linear stochastic process to a series can be constructed by computing the Fourier transform (FT) of the series, keeping unchanged the magnitudes of the Fourier coefficients, but randomizing their phases and computing the inverse FT into the time domain. Different realizations of the process are obtained using different sets of the random phases. The redundancy can be used as the nonlinear statistic for detection of nonlinearity [19]. Here we are not interested in detection of nonlinearity but in understanding what kind of changes in the data induced the differences in the CER's and whether these changes occurred in linear or nonlinear properties of the studied time series.

The marginal redundancy $\varrho^3(\tau)$ from the EEG (thick lines) and from corresponding surrogates (mean of 15 realizations – thin full lines, mean ± SD – thin dashed lines) are plotted in Figs. 5e,f. The presented results were obtained from the EEG recorded before (Fig. 5e, time zero in Figs. 5a-d) and one hour after (Fig. 5f) the ethanol administration. The decrease of $\varrho^3(\tau)$ in Fig. 5e is clearly faster than in Fig. 5f, what is reflected in related values of the CER $h^{(1)}$, which are 0.91 in time 0 and 0.66 in time 1 hour after the ethanol administration (Fig. 5b). The marginal redundancies $\varrho^3(\tau)$ from the EEG's and from corresponding surrogates in Figs. 5e,f practically coincide. It means that the dynamics of the EEG and the changes induced by alcohol are very well described by a linear Gaussian process[9] and therefore the influence of alcohol on the EEG can be successfully quantified by standard linear tools such as the spectral analysis. The CER's give the same results, because no specifically nonlinear phenomenon was observed which could induce differences between the CER's and the linear descriptors.

Different situation was observed when the CER's were applied to classification of hand tremor. Tremor is classified into physiological, essential and parkinsonian tremor by means of clinical criteria, including medical history of patients. There have been attempts to separate these tremors by time-series analysis of recordings of an acceleration of the hand tremor (see [33] and references therein). Linear, either time or frequency domain methods failed to separate the types of the tremors. Timmer et al. [33] demonstrated that the distinction of the tremors, based only on normalized time series of the acceleration of a stretched hand, was possible, using a two-step strategy: First, the physiological tremor was separated from the pathological (essential and parkinsonian) tremors using statistics for distinguishing linear and nonlinear processes. Then, the essential and parkinsonian tremors were separated using statistics that characterize nonlinear properties of time series, namely time reversal invariance and asymmetry of decay of autocorrelation functions [33].

In Fig. 6e we present results from a part (4 series of physiological, 5 series of essential, and 5 series of parkinsonian tremors) of the tremor data described in [33]. The CER's $h^{(1)}$ were computed from $\varrho^3(\tau)$ ($Q = 4$, series length 8192 samples, sampling frequency 300 Hz) using $\tau_0 = 0$ and $\tau_{max} = 1024$ samples, which satisfies the condition $\varrho^3(\tau \geq \tau_{max}) \approx 0$ for all compared series, required in Sec. 3, however, also $\tau_{max} = 256$ gives comparable results. Analysis of variance confirmed significant distinction of the three groups by the related p-value equal to 0.0005. The p-values for pairwise t-tests are: 0.042 for physiological-essential, 0.022 for physiological-parkinsonian, and 0.036 for essential-parkinsonian tremors. Thus the three types of the tremors were separated on one scale given by the easily computable CER $h^{(1)}$. Considering, however, that only a small sample of the datasets was processed, the presented result should be considered as preliminary. Larger database of tremor recordings and procedures of discriminant analysis should be used to establish efficacy of the CER's in tremor diagnostics.

---

[8] Also, preservation of histogram is usually required. A histogram transformation used for this purpose is described in Ref. [19] and references within.

[9] More exactly, a linear Gaussian model can mimic the dynamics of the EEG very well, in particular, linear description of this pharmaco-EEG study is sufficient, although generally linear processes are not able to explain *all* properties of the EEG. Even in this case the test statistic refuses the null hypothesis of linearity, which is not apparent from Figs. 5e,f, however, detected nonlinearity does not play any important role here, i.e., the studied changes occurred in linear properties of the EEG. Detailed study of nonlinearity in normal human EEG can be found in Ref. [32].

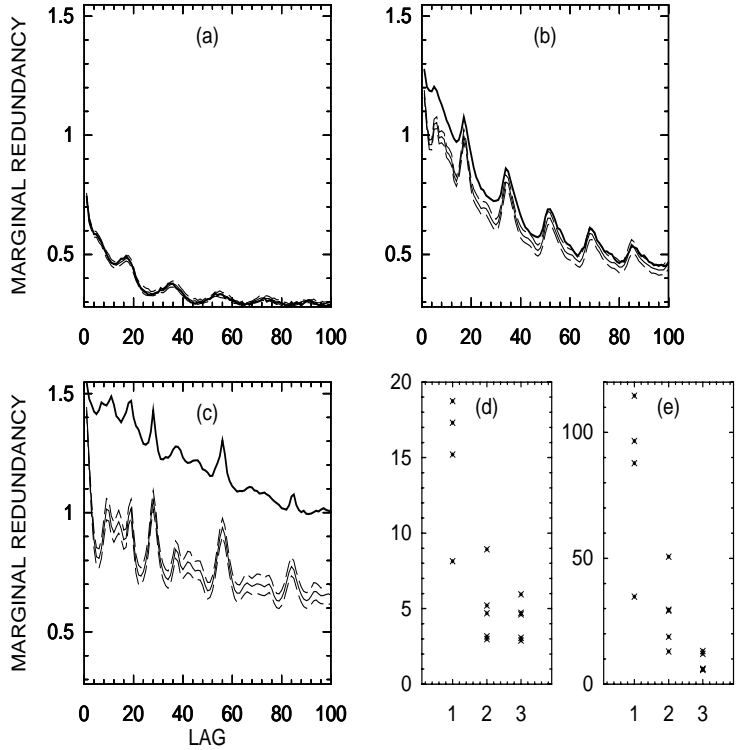

Figure 6: (a,b,c) Marginal redundancies for (a) physiological, (b) essential and (c) parkinsonian tremor (tremor data – thick lines), and for corresponding isospectral surrogate data (thin full lines – mean of a set of 15 realizations of the surrogates, thin dashed lines – mean $\pm$ SD), $N = 8,192$, $Q = 4$, $n = 3$, lags 1 – 100 samples. (d) The "linear entropy rates" – LER's, (e) the coarse-grained entropy rates $h^{(1)}$, $N = 8,192$, $Q = 4$, $n = 3$, $\tau_0 = 0$, $\tau_{max} = 1024$. Individual values are plotted for the 3 groups (abscissa): 1 = physiological (4 measurements), 2 = essential (5 measurements, 2 overlapping), 3 = parkinsonian tremor (5 measurements, several overlaps).

Gantert et al. [34] have classified the tremors as different kinds of dynamical processes: the physiological tremor was described as a linear stochastic process, the essential tremor as nonlinear and stochastic, and the parkinsonian tremor was classified as nonlinear and deterministic – chaotic[10]. One physiological phenomenon can occur in several qualitatively different dynamical realizations. This could be the reason why the previous attempts to separate the three tremors, using one measure, failed. We have tested, however, the possibility to distinguish the tremors by a linear equivalent to the CER $h^{(1)}$. We have defined the "linear entropy rate" – LER exactly by the same way as the CER $h^{(1)}$, however, absolute value of the autocorrelation function was used instead of the marginal redundancy $\varrho^n(\tau)$. The results, presented in Fig. 6d, show that the distinction by the LER of the physiological tremor from the pathological ones is similar like the results of the CER's, however, the LER failed to distinguish the essential and parkinsonian tremors.

In order to understand the results, we again present the plots of the marginal redundancy $\varrho^3(\tau)$ for the different types[11] of tremors and for corresponding surrogate data (6a – physiological, 6b – essential, 6c – parkinsonian tremor, thick lines – tremor data, thin lines – surrogates: mean of a set of 15 realizations of the

---

[10] Recent investigations [35] revealed, however, that the classification of the parkinsonian tremor as a deterministic chaotic process had been incorrect — low correlation dimension estimates reported in [34] are now considered artifacts of an improper choice of time delays used in constructions of embeddings. The parkinsonian tremor should be probably classified as a nonlinear stochastic process.

[11] The three examples (Figs. 6a-c) are typical in the sense that the results obtained from other recordings in a particular group are not qualitatively different from those presented. The only exception is one recording of the physiological tremor, $\varrho^n(\tau)$ of which resembles more the results of the essential tremors, and, consequently, $h^{(1)}$ for this recording (the lowest $h^{(1)}$ among the physiological tremors) lies inside the range of $h^{(1)}$ of the essential tremors.

surrogates – thin full lines, mean ± SD – thin dashed lines). As expected, the values of $h^{(1)}$ reflect the rates of decrease of $\varrho^3(\tau)$ of the tremors (thick lines). In the example of the physiological tremor, $\varrho^3(\tau)$ decreases with the highest rate and no differences between $\varrho^3(\tau)$ from the tremor data and $\varrho^3(\tau)$ from its surrogates are apparent, consistently with the classification of the physiological tremor as a linear stochastic process. In both cases of the pathological tremors the differences between the data and the surrogates are statistically significant, however, these differences are much larger in the case of the parkinsonian tremor than in the case of the essential tremor. The latter can be characterized as weakly nonlinear, the former as a (strongly) nonlinear process. Note that the rates of decrease of $\varrho^3(\tau)$ computed from the surrogates (thin lines) of the two pathological tremors (Figs. 6b,c) are very similar. If the entropy rates were evaluated considering only linear properties of the tremors, i.e., the linear LER's were used, or, the CER's $h^{(1)}$ were computed from the surrogates, the essential and parkinsonian tremors could be indistinguishable (Fig. 6d). Thus a nonlinear approach is necessary to distinguish the two pathological tremors, while the physiological tremor can be identified and described by linear tools. The three types of tremor have apparently different dynamical properties, however, they all possess positive entropy rates, which provide the principal possibility to compare the tremors using the one scale – the CER's, independently of linearity or nonlinearity, determinism or stochasticity of underlying processes.

# 6  Conclusion

We have introduced the coarse-grained entropy rates (CER's), quantities which are suitable for classification of experimental time series. The classification provided by the CER's, based on relative quantification of irregularity and predictability of the series (and the underlying processes), is related to the classification given by the exact entropy rates (the Kolmogorov-Sinai entropies in the case of chaotic systems), which, however, cannot be computed in majority of experimental applications. The CER's are much more robust with respect to noise than the "fine-grained" measures like dimensions and Lyapunov exponents. Considering that the entropy rates can be defined for both deterministic and stochastic processes, the application of the CER's is meaningful irrespectively of the origin of the data[12]. However, when the data can be described by a Gaussian process, then all dynamical information about a process is contained in its spectrum and the CER's cannot bring more information than the spectral analysis, as it was demonstrated in the above example from the pharmaco-EEG study. On the other hand, there are many real-world problems in which nonlinearity (either deterministic or stochastic) plays an important role and the application of the CER's can bring complementary information not detectable by standard linear tools, while applications of chaos-based measures (dimensions, Lyapunov exponents, Kolmogorov-Sinai entropy) can be disqualified by high dimensionality or stochasticity of processes under study, or even by quality of processed data (noise, finite precision).

## Acknowledgements

The author would like to thank I. David for the pharmaco-EEG data and G. Deuschl and J. Timmer for the tremor data. Special thanks are due to J. Klaschka, D. Prichard, J. Theiler, D. Kaplan and Z. Kowalik.

The author was supported by the International Research Fellowship F05 TWO4757 ICP from the National Institutes of Health, the Fogarty International Center, and also by grants to the Santa Fe Institute, including core funding from the John D. and Catherine T. MacArthur Foundation, the National Science Foundation (PHY-8714918), and the U.S. Department of Energy (ER-FG05-88ER25054), and in part by the Grant Agency of the Academy of Sciences of the Czech Republic (grant No. 230404).

---

[12] Of course, with exemption of regular deterministic processes with zero entropy rate, which can be easily identified from the shape of the marginal redundancy curve plotted as a function of time lag $\tau$ [15].